\documentclass{article}
\usepackage{setspace}
\usepackage{fullpage}
\usepackage{scicite}
\usepackage{amsfonts}
\usepackage{amssymb,enumerate}
\usepackage{amsmath} 
\usepackage{rotating}
\usepackage{fancyvrb}

\title{Yahtzee\texttrademark: An Anonymized Group Level Matching Procedure}

\author{Jason J. Jones$^1$ \and Robert M. Bond$^2$ \and Christopher J. Fariss$^2$ \and Jaime E. Settle$^2$ \and Adam Kramer$^3$ \and Cameron Marlow$^3$ \and James H. Fowler$^{2,4}$ \\ 
\\
\small {$^1$ Psychology Department, University of California, San Diego, La Jolla, CA 92093, USA} \\
\small {$^2$ Political Science Department, University of California, San Diego, La Jolla, CA 92093, USA} \\
\small {$^3$ Data/Science, Facebook, Inc., Palo Alto, CA 94304, USA} \\
\small {$^4$ Medical Genetics Division, University of California, San Diego, La Jolla, CA 92093, USA}
}
\date{}

\begin{document}
\pagenumbering{roman} 
\maketitle
\begin{abstract}
Researchers often face the problem of needing to protect the privacy of subjects while also needing to integrate data that contains personal information from diverse data sources in order to conduct their research. The advent of computational social science and the enormous amount of data about people that is being collected makes protecting the privacy of research subjects evermore important. However, strict privacy procedures can make joining diverse sources of data that contain information about specific individual behaviors difficult. In this paper we present a procedure to keep information about specific individuals from being ``leaked`` or shared in either direction between two sources of data. To achieve this goal, we randomly assign individuals to anonymous groups before combining the anonymized information between the two sources of data. We refer to this method as the Yahtzee\texttrademark procedure, and show that it performs as expected theoretically when we apply it to data from Facebook and public voter records.

\end{abstract}
\clearpage
\doublespace
\pagenumbering{arabic}

\section{Introduction}
Computational social science is an emergent field of inquiry that promises to revolutionize the way we study and understand human behavior \cite{Cioffi-Revilla:2010, Lazeretal:2009}. Unfortunately, obstacles exist that hamper analysis of large-scale dynamic data sets that are now available \cite{Lazeretal:2009}. One problem is that companies such as Google, Facebook, and cell phone companies, are often reluctant to share data obtained from their clients with external researchers. When they do allow access to the data, it is often in an aggregated or anonymized format designed to protect the identities of their users. While this approach has led to a variety of collaborative research projects \cite{GonzalezHidalgoBarabasi:2008}, once identifying information is removed from the data it cannot be combined with other data sources, limiting the type and scope of research that can be performed with such data. This paper presents a procedure designed to address this issue by keeping information about specific individuals private while still allowing researchers to combine multiple sources of information in such a way that inferences may be drawn about the relationships between variables across sets of data.

We developed this method in order to join information from public voting records with data we are analyzing through a collaborative research project with Facebook \cite{Bondetal:2011}. Although our human subjects protocol approved by the Institutional Review Board at the University of California, San Diego, allows us to perform one-to-one matching of Facebook data and voter records, Facebook asked us to design a procedure that would better protect the privacy of its users. So, in order to study the voting behavior of users, it was necessary to devise a process for matching users to their publicly available voting records without identifying the behavior of specific users. 

In this process, we wanted to be sure that information about specific individuals was not ``leaked'' or shared in either direction. Our goal was to avoid connecting any specific Facebook user's voting behavior to Facebook's database of information about a given user. To achieve this goal, we devised a group-level matching procedure that computed the relationship between site usage behavior and voting behavior based on repeated random assignment of individuals to groups, as we describe below.

\section{Group-Level Anonymous Matching}
In order to begin the procedure, we needed a unique identifier for each individual that could be obtained using information in either the Facebook data or the voting record data. We used first name, last name, and date of birth (dropping all instances that had duplicates) for the voting records of one state to generate an encrypted one-way hash (we used the last 7 digits of SHA-256). 

A ``match'' is defined as a row in both files that has the same values across files for ALL of these columns: \texttt{first\_name}, \texttt{last\_name}, \texttt{birth\_day}, \texttt{birth\_month}, \texttt{birth\_year}. Duplicate rows in both files are thrown out before any matching begins.  Approximately 0.5\% of Facebook users were dropped due to duplication and also approximately 0.5\% of voters from the voter files were dropped due to duplication.

The hash takes a string of text (in this case the first name, last name, and date of birth of the user) and returns a 256 bit string that can be treated as a numeric value. We divided this value by $N/g$, where $N$ is the number of individuals in the public voter record and $g$=5 was chosen arbitrarily, and recorded the remainder as the group ID. This procedure results in groups of various sizes, but on average, groups will be of size $g$. Sampling variation caused some groups to have more or fewer than $g$ people in them; these groups were discarded. We then recorded the number of voters in each retained group, along with its group ID. Groups could have between 0 and $g$ voters in them, and the mean number of voters per group approximated the average turnout in the state, with a very small amount of sampling variation due to groups with more or less than $g$ individuals being discarded.

Facebook hashed the user record data (with the same sequence of random seeds), using first name, last name, and date of birth (also dropping duplicates) for users who logged in from the state in question on Election Day.  The Facebook data does not have \texttt{first\_name} and \texttt{last\_name} columns, but it does have a \texttt{name} column that contains the name provided by the user at time of registration. We defined \texttt{first\_name} as the first token in \texttt{name} and \texttt{last\_name} as the last token in``name''. This works well because most people enter their name such as ``First M. Last''. However, it does not work if the name is entered as ``The Illustrious First M. Last, Esquire,'' which occasionally happens online.

Facebook then divided the hash value derived from each name and birthdate by $N/g$ in order to create a group ID (note that $N$ still represents the number of individuals in the \emph{public voter record}, not the number of individuals Facebook recorded as logging in from that state). This procedure is identical to the procedure used to create group IDs using public voting records. Therefore, individuals with the same first name, last name, and date of birth in both the public voting records and Facebook's data were assigned the same group ID. That is, this procedure guaranteed that any and all Facebook users who were also registered voters would be assigned the same group ID in both sets of data. However, because there was no guarantee that a given registered voter would also be a Facebook user nor that a given Facebook user would be registered to vote (and therefore in the voter record), this procedure prevented identification of a specific Facebook user's behavior. The proportion of truly matched Facebook users in any group was unknown and could range from 0\% to 100\%.

Using the state voting data, the number of registered voters in each group who did vote (some number between 0 and $g$) in 2010 was calculated. That number was recorded and assigned to each Facebook user with the same group ID. Importantly, a Facebook user was assigned to a group whether or not they were on the registration list. The group may have had any number of voters (0 to $g$). So, in a given instance a user who was not on the registration list may be assigned to a group in which any fraction of those on the registration list voted. This feature of the procedure ensures that we cannot be certain that a particular Facebook user registered or voted based on the turnout value of their assigned group.

We repeated this procedure $m$ times, re-hashing using different seeds (and thus re-grouping individuals), and assigning an additional value to each user after every round. (This gave rise to our nickname for the procedure, ``Yahtzee\texttrademark'', which refers to the idea of metaphorically re-rolling the dice on each iteration to place users in new groups.) Thus, each user was assigned a distribution of $m$ values, from the set $y \in \{0,1, \ldots, g\}$. If $g>1$ then it is not possible to infer with certainty the voting behavior of any users, or even their registration status. However, given the distribution of these values, each additional draw provides more information and we can set $m$ such that we have enough observations per person to classify individuals on Facebook as matched voters or matched abstainers with a minimum pre-determined level of measurement error (we chose a value of 5\%).

To see why, notice that Facebook users who were not registered to vote would have an effectively random classification in every round. They are also randomly assigned to groups that have a random number of voters and abstainers in them. Therefore, if $p$ is equal to the turnout rate, then the probability that the $j$th draw for user $i$ is equal to $y$ can be determined from a binomial distribution:
\begin{align}
Pr(y_{ij}=y) = {g \choose y} p^{y} (1-p)^{(g-y)}.
\end{align}
Meanwhile, users who were registered to vote would be somewhat more likely to have the correct classification (voter or abstainer). Given that the user was on the registration list, their presence in their own group in each draw skews the distribution of their own draws (toward $g$ for voters and toward 0 for abstainers). 

Specifically, if a Facebook record does match a voter record, then its own contribution to the total number of voters in the group is always 1, and since the other $g-1$ group members are randomly assigned, the probability that a draw is equal to $y$ is
\begin{align}
Pr(y_{ij}=y) = {g-1 \choose y-1} p^{y-1} (1-p)^{(g-y-1)}.
\end{align}
By the same reasoning, if a Facebook record matches an abstainer record, then its own contribution to the total number of voters in the group is always 0, and the probability that a draw is equal to $y$ is simply
\begin{align}
Pr(y_{ij}=y) = {g-1 \choose y} p^{y} (1-p)^{(g-1-y)}.
\end{align}

Since these are independent draws, the probability of observing the set of draws $y_i$ conditional on being unregistered, a voter, or an abstainer is
\begin{align}
Pr(y_i|i\text{ is unregistered}) &= \prod_{j=1}^m{{g \choose y_{ij}} p^{y_{ij}} (1-p)^{(g-y_{ij})}} \\
Pr(y_i|i\text{ is a voter}) &= \prod_{j=1}^m{{g-1 \choose y_{ij}-1} p^{y_{ij}-1} (1-p)^{(g-y_{ij}-1)}} \\
Pr(y_i|i\text{ is an abstainer}) &= \prod_{j=1}^m{{g-1 \choose y_{ij}} p^{y_{ij}} (1-p)^{(g-y_{ij}-1)}}.
\end{align}

We can use these probabilities to classify individuals, assigning each to the classification that maximizes the likelihood of observing $y_i$. For improved efficiency we transform the equations to log likelihoods, and we use simulations to estimate the number of values needed per record ($m$) to generate a specific classification error. Simulation code (written in R) is provided below.

For any application, we must select two values of $m$ for each set of records that we wish to match in order to balance the rate of false voters and false abstainers. This is because the overall turnout rate determines which behavior takes fewer observations to distinguish from average behavior. If most people abstained, it will take fewer observations to identify groups where users likely voted, and vice versa.  We therefore must make additional draws for individuals classified as belonging to the more common group.  To achieve balanced rates we select two values: $m_1$ is the number of draws necessary to reach the desired level of accuracy for the less common behavior and $m_2$ is the number of additional draws necessary to reach the desired level of accuracy for all individuals classified with the more common behavior after $m_1$ draws.  

Choosing $m_1$ and $m_2$ requires knowledge of the aggregate turnout rate, which was computed directly from the voter record. It also requires knowledge of the match rate (the probability a given Facebook record can be matched to a specific voter record). For each state, Facebook estimated the match rate by drawing 1000 records at random from their database, and counting the number of matches with a list of the names and birthdates that were available in the voter record. No individual match was recorded: Only the aggregate match rate was stored, and all other information was discarded. 

\begin{figure}[h]
\begin{center}$
\begin{array}{cc}
\includegraphics[width=2.75in]{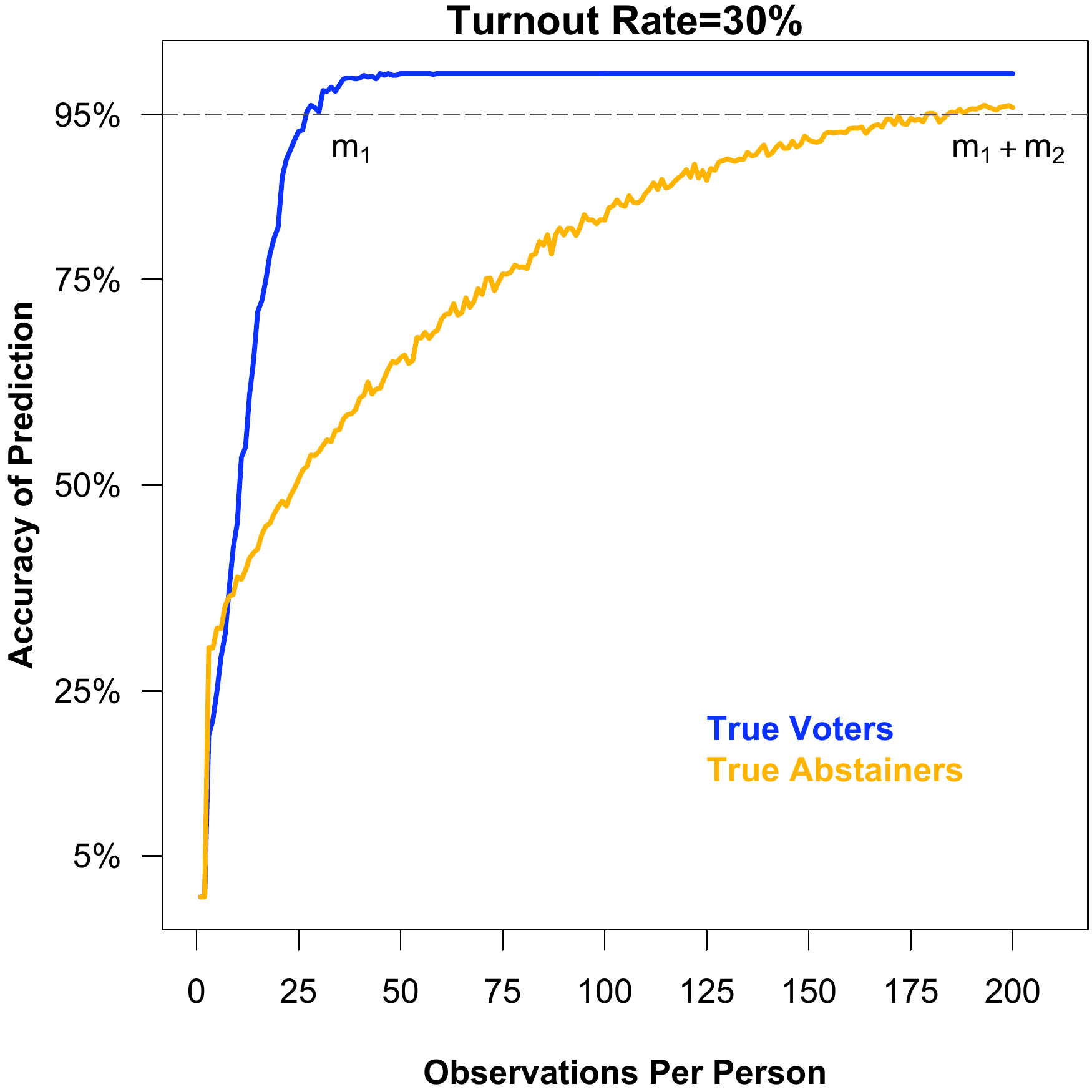} &
\includegraphics[width=2.75in]{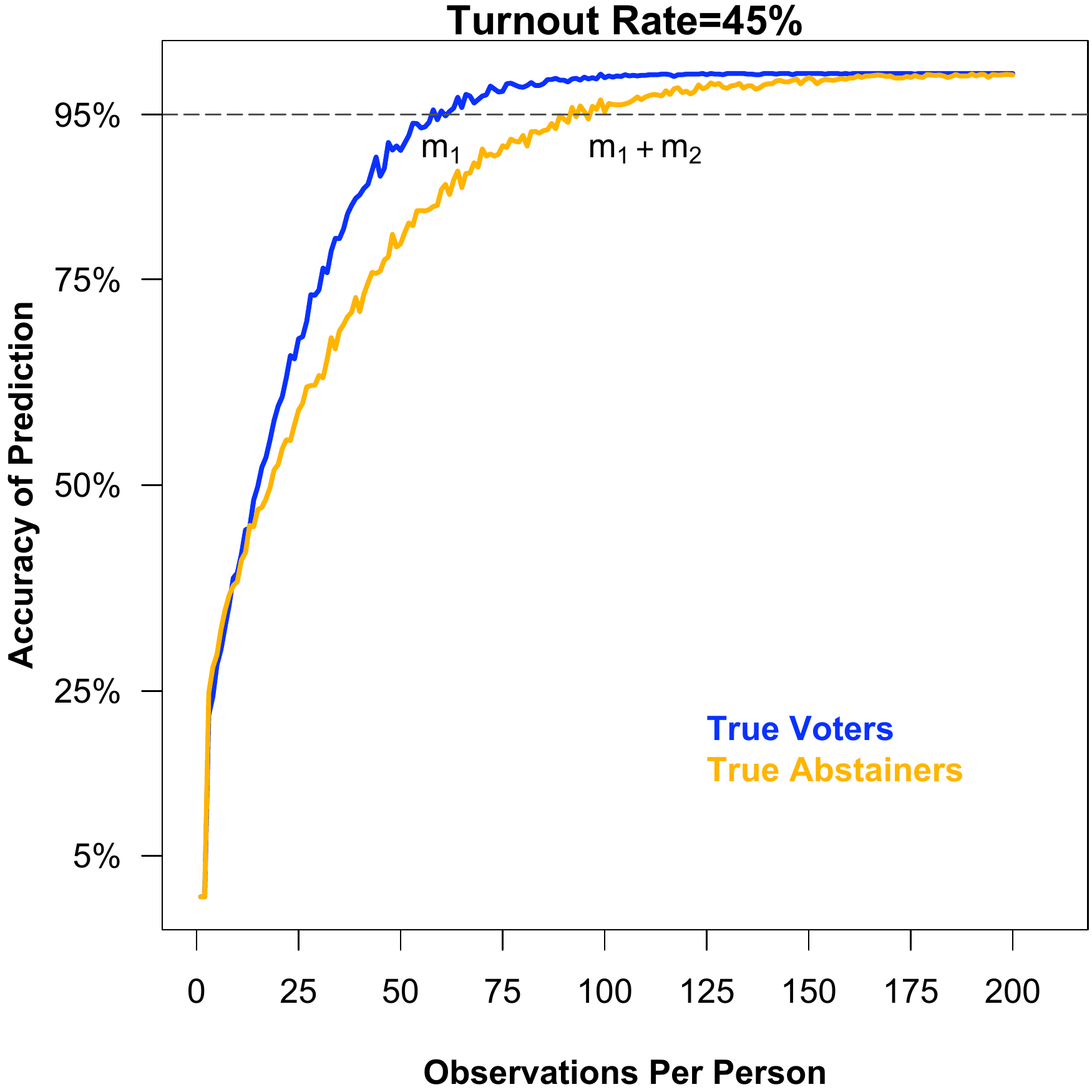} \\
\\
\includegraphics[width=2.75in]{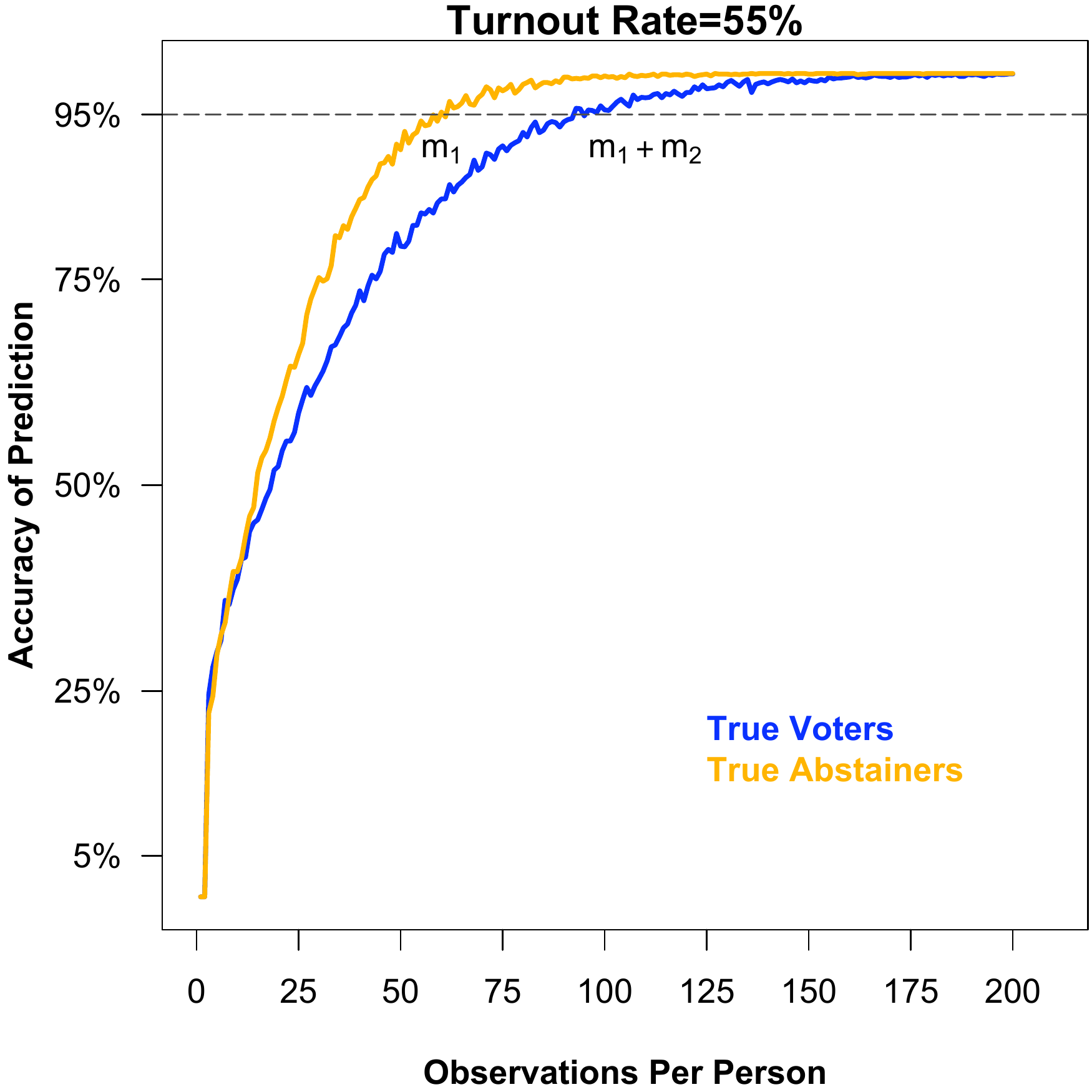} &
\includegraphics[width=2.75in]{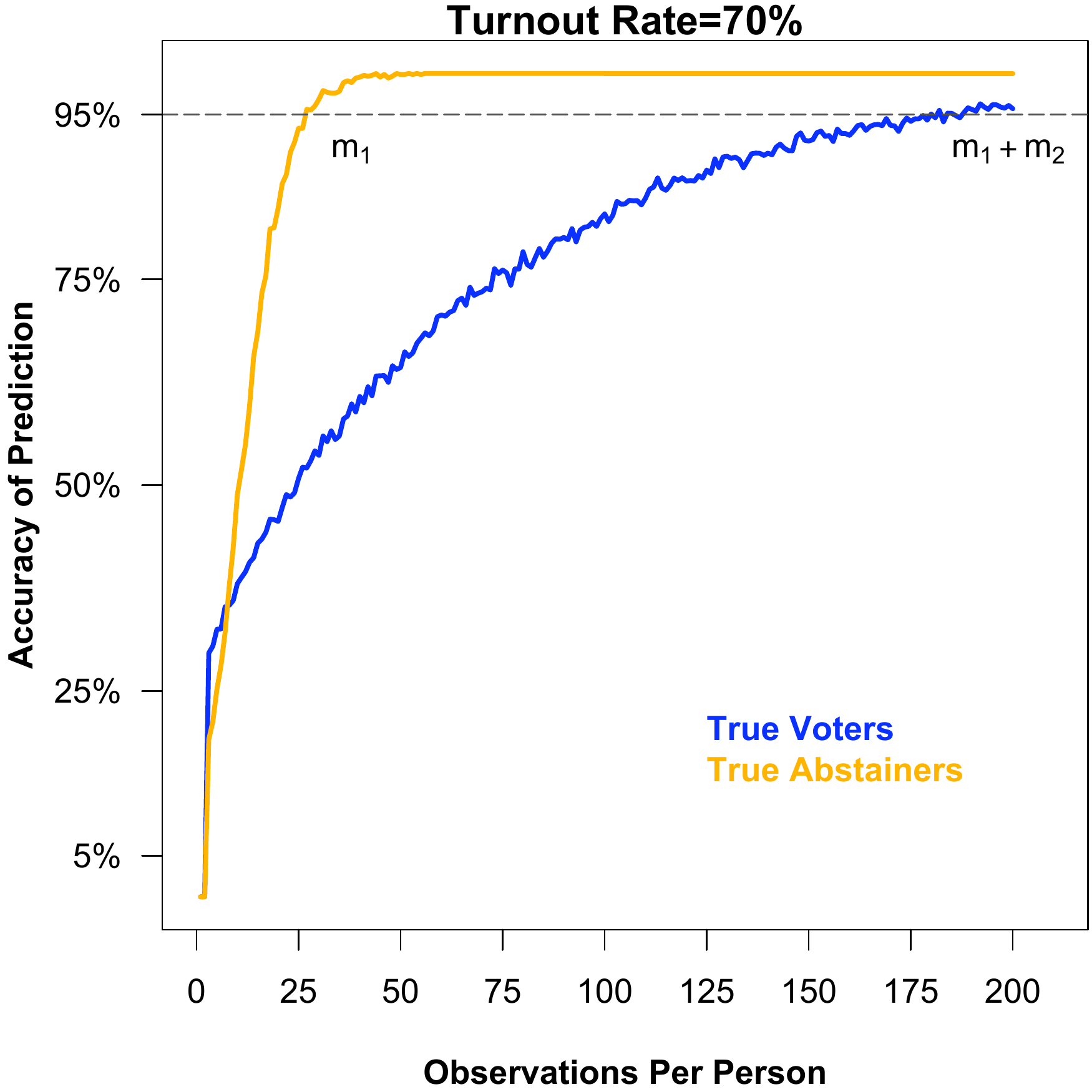} \\
\end{array}$
\end{center}
\caption{The proportion of correct predictions for participation rates of 30\%, 45\%, 55\%, and 70\% (the match rate is held constant at 30\% in all four figures) from a simulation of the matching procedure. The dark line represents the accuracy rate for true participators. The light line represents the accuracy rate for true abstainers. accuracy increases for both categories as observations for each individual are obtained from the Yahtzee\texttrademark procedure. Note that the less common of the two behaviors requires fewer observations for classification than the more common behavior. $m_1$ is the number of observations per person necessary to achieve a given level of accuracy for the less common behavior and $m_1 + m_2$ is the number of observations necessary to achieve a given level of accuracy for the more common behavior.}
\label{PrecisionRplot.pdf}
\end{figure}

In order to test the method we simulated the matching procedure using a set match rate that approximates what we observed in the 13 states that we used to match voter data (30\%).  We also set the turnout level to match each state in order to assess the prediction error associated with a given number of draws. The results of these simulations are summarized by Figure \ref{PrecisionRplot.pdf}. The simulations show that the less common behavior takes fewer observations to classify individuals with a given level of confidence, and that as the turnout moves away from 50\%, more observations ($m_2$) are needed to reach the level of confidence of the less common behavior.

It is important to note, once again, that the procedure only gives us estimates of the probability that any given Facebook user is on the registration list and estimates of their voting behavior. We can not be certain whether or not a user is on the list, has voted, or has abstained from these draws. In fact, it is possible that a voter will be classified as an abstainer or the reverse. The number of draws is chosen such that classifications of this type are unlikely, but still possible. Using $m_1$ and $m_2$ we are able to set the measurement error level (in our case 95\%) that we determined would be an appropriate balance between capability of inference and protection of privacy of users. In other research applications a higher or lower level of measurement error may be desired, which can easily be achieved by adjusting $m_1$ and $m_2$ accordingly.

\section{Validation}
To choose which states to validate, we identified those that provided (for research purposes) first names, last names, and full birth dates in publicly available voting records. From these, we chose a set that minimized cost per population. Of these states, the cost of voting records varied from \$0 to \$1500 per state. Costs were even greater in other states. We excluded records from Texas because they systematically excluded some individuals from their voting records (specifically, they did not report on the voting behavior of people that had abstained in the four prior elections). The resulting list of states included Arkansas, California, Connecticut, Florida, Kansas, Kentucky, Missouri, Nevada, New Jersey, New York, Oklahoma, Pennsylvania, and Rhode Island, and yielded 6,338,882 records of voters and abstainers. This value reflects the fact that we obtained about 1/3 of all voter records in the U.S., and of those, about 1/3 matched to the 61 million users who logged in on Election Day.

To validate the Yahtzee\texttrademark process, we compared its classifications for a small set of randomly chosen records for each state to true voting behavior. Table \ref{redraw} contains the m1 and m2 values for each state. These values were chosen by adjusting the $m_1$ and $m_2$ values in the simulation code in the Appendix until values yielding approximately 95\% accuracy were found. As seen in the table, the values for $m_1$ and $m_2$ vary considerably. The variation in $m_1$ is due in large part to variation in the match rate. A lower match rate requires more draws overall in order to distinguish those on the registration list from those not on the registration list. Variation in $m_2$ is primarily due to variation in the turnout rate in the states. States that have a turnout rate near 50\% (such as Florida and Kansas) take few extra observations to distinguish the less common behavior from those who are assigned values at random, while states with a turnout rate far from 50\% (such as Arkansas and New Jersey) require many extra draws to make such distinctions.

Table \ref{truth} shows conditional probabilities generated from truth tables for the Yahtzee\texttrademark classifier results. For each state, 1000 Facebook user records were chosen at random. Each was given a classification based on the Yahtzee\texttrademark process. The truth tables contained the frequency of each classification that was assigned to each true behavior. This information was used to calculate the classification accuracy in the categories of interest (voter or abstainer), which are displayed for each state in Table \ref{truth}.  We also calculated the 95\% confidence interval for a null hypothesis that the prediction is correct 95\% of the time (based on an assumption the successes are binomially distributed from the same number of draws observed).  Note that nearly all of the confidence intervals contain the observed data, suggesting that deviations from 95\% accuracy are due to sampling variation. 

Researchers interested in this procedure have the ability to increase or decrease the accuracy of the group level matching procedure by increasing or decreasing the number of observations generated for each user. At the limit (extremely high values of $m_1$ and/or $m_2$) researchers have the ability to draw enough observations that they are extremely confident about the true behavior of users, but because of the group-level matching nature of the procedure, they will still never be 100\% certain of an individual's behavior.

In addition to estimates of the voting behavior of individuals, the procedure gives us estimates of the probability that an individual is on the voting record at all. As Table \ref{truth} shows, there is approximately a 99\% chance that when we do not find a match for a user that there was not a match for that individual on that state's voting record.  However, we required a perfect match for first name, last name, and birthdate.  Thus, while we might be nearly certain that there was no match, the presence of nicknames, variation in reported birth date, and other errors in the data mean that unmatched users might actually be in the voter record.  So this confidence level is an upper bound and the probability that a user is not in the record given that we classified them as not matched is probably lower than 99\%.  This means that interpretations of analyses based on the unmatched classification should be careful to describe them as measuring the match rate rather than measuring the exact likelihood that a given user was in the record.  It also means that user privacy is more protected by uncertainty since there is a greater chance that they were actually in the record when the procedure classifies them as unmatched.

\begin{figure}[h]
\begin{center}$
\begin{array}{c}
\includegraphics[width=4.50in]{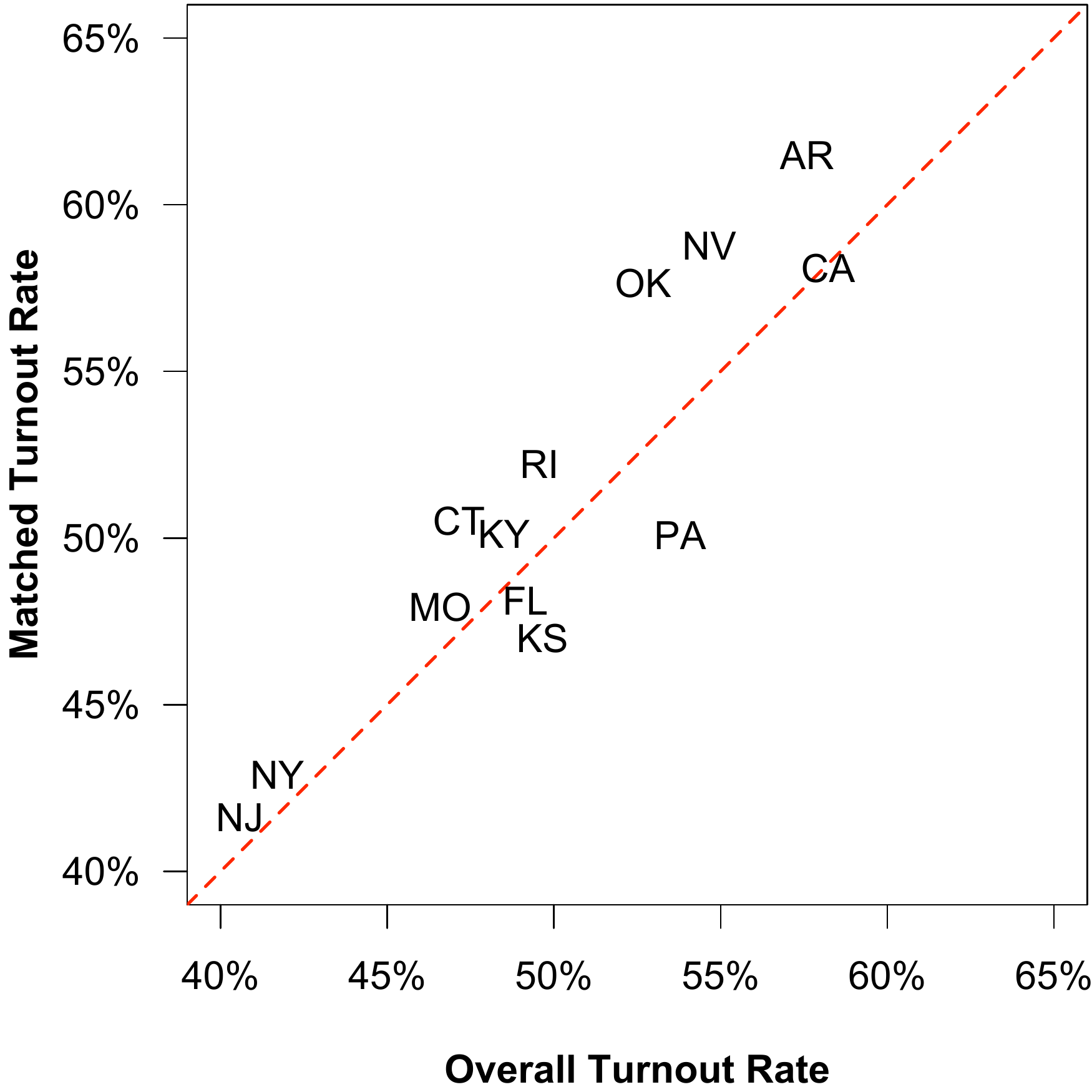} 
\end{array}$
\end{center}
\caption{The proportion of matched users who turned out to vote compared to the overall turnout rate by state. Note that the abbreviation for Kansas is repositioned slightly so that it does not overlap with the abbreviation for Florida. The results show that the Yahtzee\texttrademark procedure produces about the same overall turnout rates for each state as those shown in the voter record.} 
\label{Match_by_Overall_Rplot.pdf}
\end{figure}

Although there may be important systematic differences between users with matchable and unmatchable records, our results suggest the difference is not large.  Figure \ref{Match_by_Overall_Rplot.pdf} shows that there is a good fit between the turnout rate of matched Facebook users and the overall turnout rate of each state. This positive relationship suggests that the matching procedure is producing reliable estimates of turnout for matched users at an aggregate level.  However, the relationship is not perfect for at least two reasons. First, Facebook users within a given state are not necessarily a representative sample of that state's population. For example, we know that the age of Facebook users skews toward younger people.  Second, matched users are not necessarily representative of all Facebook users, including those who could not be matched.  For example, people who use exotic nicknames may have personality traits that also affect their willingness to vote.  Thus, while the good aggregate level fit is suggestive, we should be cautious when describing our results to explain the limitations of out-of-sample inferences that might be made using the matched data.

\begin{figure}[h]
\begin{center}$
\begin{array}{c}
\includegraphics[width=4.50in]{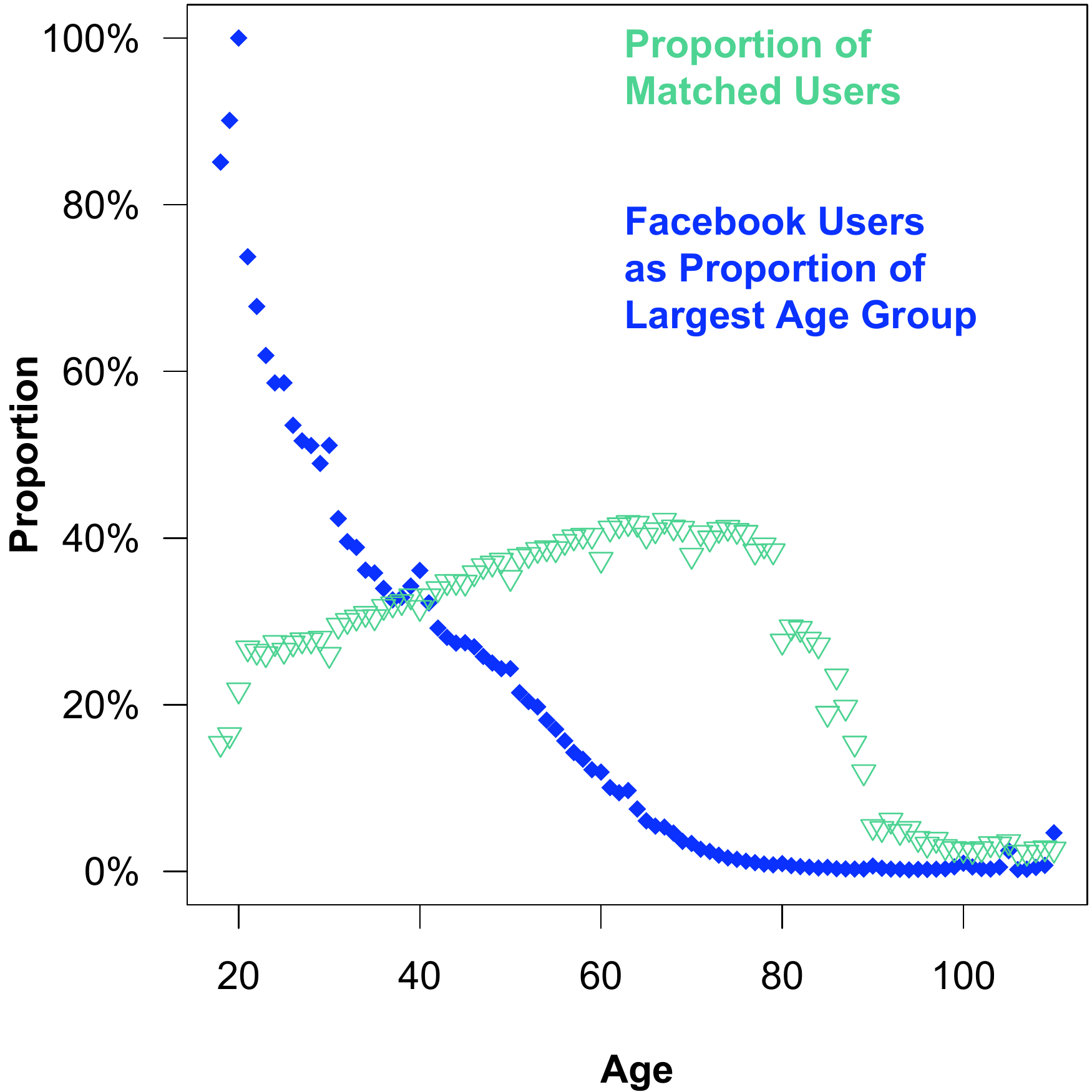} 
\end{array}$
\end{center}
\caption{The proportion of Facebook users that were matched to the validated voting record by age (triangles) and each age group's proportion of the largest age group, those 20 years of age at the time of the election (diamonds).  This figure helps to explain why match rates are lower for Facebook users who tend to be younger and more difficult to match than the average registered voter.} 
\label{ProportionRplot.pdf}
\end{figure}

About 1 in 3 users were successfully matched by the Yahtzee\texttrademark process (note that success depends on many factors, including voting eligibility, rates of registration, and so on). Although the match rate for this study is lower than the match rates in many other GOTV studies (which is usually about 50\%), this may be due to the demographic composition of Facebook.  In Figure \ref{ProportionRplot.pdf} we show how the probability of matching varies by age. There is a positive relationship between age and the probability of matching the voting record through approximately age 80 (as seen by the positive slope of the triangles). While there is a drop off in the probability of obtaining matches for users over the age of 80, it is important to note that there are very few Facebook users in this age group (as seen by the left skew of the diamonds). Younger users are also more difficult to match, likely because fewer of them are registered, and even those who are registered may be accessing Facebook from an out-of-state college. Older Facebook users are easier to match, but there are fewer of them. 

\begin{figure}[h]
\begin{center}$
\begin{array}{c}
\includegraphics[width=4.50in]{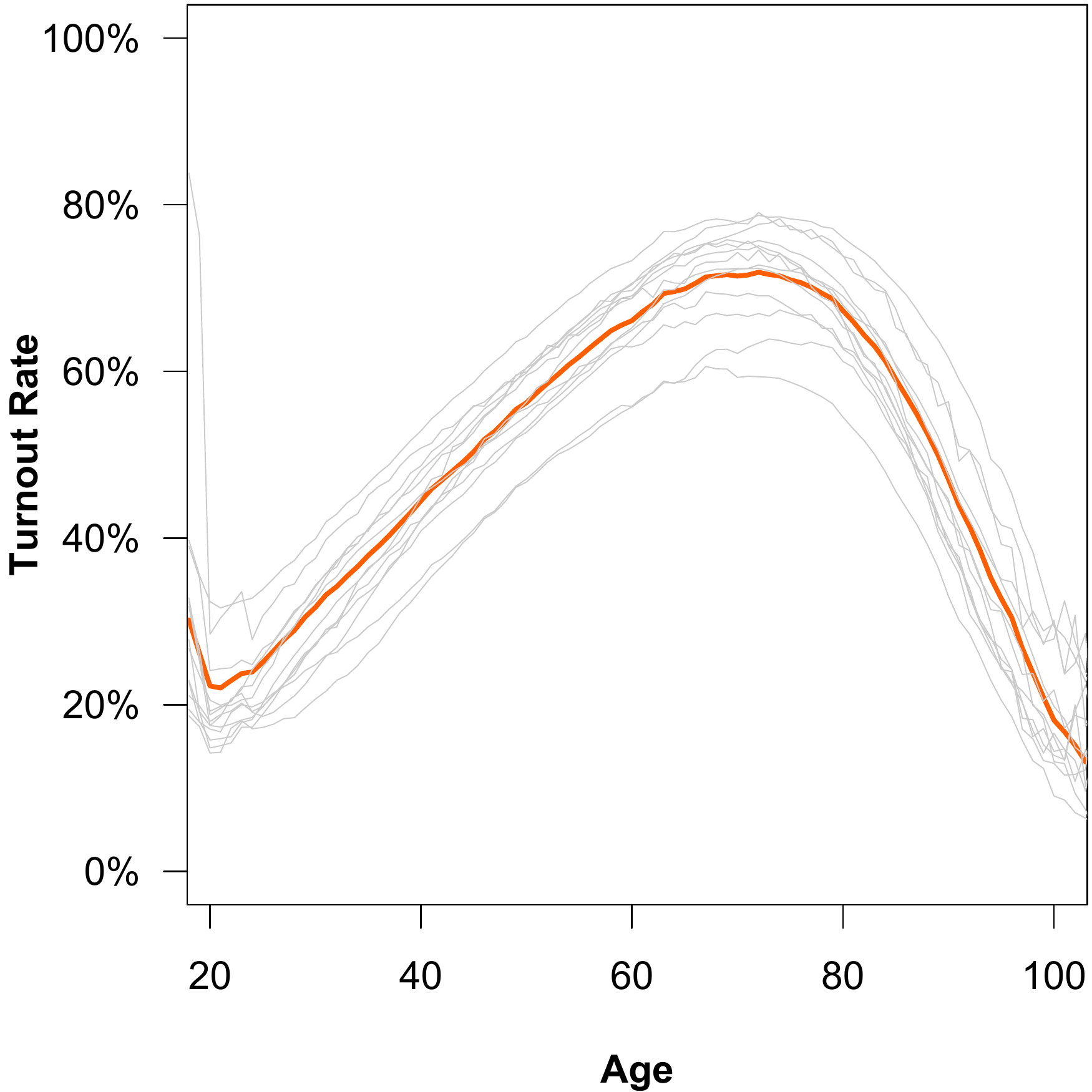} 
\end{array}$
\end{center}
\caption{The proportion of matched users who turned out to vote by age. The orange line represents the turnout rate by age of the matched sample of Facebook users. Each gray line represents the turnout rate by age of a state voter record.  The results show that users on Facebook exhibit the same pattern of turnout with respect to age as the populations in other states.} 
\label{TurnoutRateRplot.pdf}
\end{figure}

Because we know that the matched sample is not representative of the overall population by age, we assessed the turnout rate of the matched Facebook user sample as it compared to the turnout rate of each states by age. Figure \ref{TurnoutRateRplot.pdf} shows that the turnout rate goes up among older users and declines with advanced age in both the matched sample and the voter records for each state.  These results suggest that the matching procedure correctly identifies voters and abstainers and that once we control for the skew in the age distribution of the matched sample, the voting behavior of Facebook users is not very different from that of the population overall.

\section{Discussion}
Here we have introduced a method of group-level matching that allows researchers to merge two data sources while respecting the privacy of individuals in the constituent data sets. Methods like these are essential for researchers who are interested in making inferences that draw upon data about individuals, while also respecting the privacy of individuals (and the privacy policies of entities that collect such data). Many extraordinary research projects could be enhanced by joining their data with other sources of individual information. For example, one project collected 509 million twitter messages from 2.4 million individuals from 84 countries between February 2008 and January 2010 \cite{GolderMacy:2011}. Other studies include an analysis of the mood within America as function of date and time using 300 million twitter messages generated between September 2006 and August 2009 \cite{Misloveetal:2011}, an analysis of 50 million Google search queries to identify the weekly influenza level in regions of the United States \cite{Ginsbergetal:2009}, and an analysis of the application adoption patterns of 50 million Facebook users \cite{OnnelaaReed-Tsochas:2010}. In each of these studies however, a variety of additional questions could be addressed if more information about the users generating the observed data could be obtained. This information often exists in other datasets yet linking these datasets raises both technical and ethical concerns. 

To address these issues we have developed a method to anonymously match group level information. This method is different from other ``privacy preserving'' machine learning techniques in that it joins information about individuals from at least two sources \cite{LindellPinkas:2009}. ``Privacy preserving'' machine learning techniques pool similar data from multiple sources. The ``Yahtzee\texttrademark`` procedure also does not rely on a third party function to join the constituent datasets as some other procedures do \cite{FreedmanNissimPinkas:2004}. For a review of other computational methods for ensuring case level anonymity see \cite{Goldreich:2003}. The advantage of the ``Yahtzee\texttrademark`` procedure is that no party ever needs to know certain information about a specific individual because this information is anonymized prior to merging the constituent datasets. The Yahtzee\texttrademark procedure can therefore be used to preprocess a dataset before it is ever sent to another organization for analysis. Thus, multiple data collecting entities can use this procedure to anonymize their data before combining that data with other sources for joint analysis.

We used the Yahtzee\texttrademark method to match public voting records to Facebook user data.  This application allowed us to test our method on real world data, where we found that it generates the same level of uncertainty about individual records that was predicted by theory. Additionally, we found that the turnout rate of Facebook users by state strongly correlates with the overall turnout rate of all individuals in the state and Facebook users within each age group tend to vote at about the same rate as members of those age groups in the population as a whole.  These results not only suggest that the Yahtzee\texttrademark method works as expected, but also that Facebook users are very similar to the population as a whole in terms of their voting behavior.  This should be an encouraging result for a growing group of researchers who rely on internet websites such as Facebook or Amazon's Mechanical Turk in order to recruit subjects.

We live in an age in which more and more data is being collected about individuals, providing researchers with the opportunities to study phenomena at a scale never possible before and to study new relationships that were infeasible before due to the difficulty of collecting information from diverse sources about the same individuals. While the availability of this data offers exciting opportunities for new avenues of research, much of it is held by corporations that have an interest in maintaining the privacy of their users or customers. In order for researchers to conduct studies using this data, new methods will need to be invented that fit specific problems with the data. In this paper, we offer a solution to what we believe is a common problem that corporations and researchers often face, matching sensitive data to other data sources.

\clearpage
\section{Appendix}
\subsection{R Code Part 1}
\singlespace
\begin{Verbatim}[frame=single]
t<-0.45  # turnout rate
mm<-0.3  # match rate
n<-100000 # total sample
g<-5   # group size
m<-56   # observations per person
m2<-27  # additional observations per person to eliminate asymmetry

n0<-round(mm*(1-t)*n) # number of matched abstainers
n1<-round(mm*t*n)   # number of matched voters
n2<-n-n0-n1      # number of mismatched

# 0 = matched abstainer, 1 = matched voter, 2 = mismatched
type<-c(rep(0,n0),rep(1,n1),rep(2,n2)) 

# generate matched abstainer observations
y0<-matrix(rbinom(n0*m,g-1,t),ncol=m)

# generate matched voter observations
y1<-matrix(rbinom(n1*m,g-1,t)+1,ncol=m)

# generate mismatched observations
y2<-matrix(rbinom(n2*m,g,t),ncol=m)

# join groups
y<-rbind(y0,y1,y2)

# initialize prediction vector
pred_type<-rep(NA,n)

# loop through each person to predict type and probabilities
for(i in 1:n) {
 pred_type[i]<-which.max(c(sum(dbinom(y[i,], g-1,t, log=T)),
              sum(dbinom(y[i,]-1,g-1,t, log=T)),
              sum(dbinom(y[i,], g, t, log=T))))-1
}
\end{Verbatim}
\clearpage

\clearpage

\subsection{R Code Part 2}
\begin{Verbatim}[frame=single]
# redraw the less frequent types to increase accuracy
if(m2>0) {

 # identify redo set
 redo_set<-which(pred_type==ifelse(t<0.5,0,1))
 
 # generate more matched abstainer observations
 yy0<-matrix(rbinom(n0*m2,g-1,t),ncol=m2)

 # generate more matched voter observations
 yy1<-matrix(rbinom(n1*m2,g-1,t)+1,ncol=m2)

 # generate more mismatched observations
 yy2<-matrix(rbinom(n2*m2,g,t),ncol=m2)

 # join groups
 yy<-rbind(yy0,yy1,yy2)

 # loop through each person in redo set to predict type
 for(i in redo_set) {
 pred_type[i]<-which.max(c(sum(dbinom(c(y[i,],yy[i,]), g-1,t, log=T)),
              sum(dbinom(c(y[i,],yy[i,])-1,g-1,t, log=T)),
              sum(dbinom(c(y[i,],yy[i,]), g, t, log=T))))-1
 }
}

# probability predicted matched abstainer is correctly identified
sum(pred_type==0&type==0)/sum(pred_type==0)

# probability predicted matched voter is correctly identified
sum(pred_type==1&type==1)/sum(pred_type==1)
\end{Verbatim}

\clearpage

\clearpage

\begin{table}[ht]
\begin{center}
\begin{tabular}{|lccc|}
 \hline
 State & $m_1$ & $m_2$ & Common Type \\
 \hline
 \hline
Arkansas & 55 & 45 &  Voters \\
California & 50 & 50 &  Voters \\
Connecticut & 65 & 10 &  Voters \\
Florida & 75 &  0 & Abstainers \\
Kansas & 75 &  0 & Abstainers\\
Kentucky & 60 &  5 &  Voters \\
Missouri & 70 & 20 &  Abstainers \\
New Jersey & 60 & 65 &  Abstainers \\
Nevada & 65 & 25 &  Voters \\
New York & 55 & 50 &  Abstainers \\
Oklahoma & 65 & 15 &  Abstainers \\
Pennsylvania & 65 & 15 &  Voters \\
Rhode Island & 75 &  5 &  Abstainers \\
  \hline
\end{tabular}
\end{center}
\caption{$m_1$ is the number of draws necessary to reach the desired level of accuracy for the more common behavioral type and $m_2$ is the number of additional draws necessary to reach the desired level of accuracy for the less common behavioral type.}
\label{redraw}
\end{table}

\begin{table}[ht]
\begin{center}
\begin{tabular}{|clcc|}
  \hline
 State & & Pr() & 95\% CI \\ 
 \hline
 \hline
			& Pr(Abs$|$Class=Abs) & 0.949 & [0.908, 0.990] \\ 
 Arkansas		& Pr(Vot$|$Class=Vot) & 0.957 & [0.913, 0.981] \\ 
 			& Pr(NM$|$Class=NM) & 0.988 &  \\ 
 \hline
 			& Pr(Abs$|$Class=Abs) & 0.932 & [0.912, 0.980] \\ 
 California	& Pr(Vot$|$Class=Vot) & 0.951 & [0.919, 0.978] \\ 
 			& Pr(NM$|$Class=NM) & 0.987 &  \\ 
 \hline
 			& Pr(Abs$|$Class=Abs) & 0.888 & [0.916, 0.978] \\ 
 Connecticut	& Pr(Vot$|$Class=Vot) & 0.988 & [0.912, 0.981] \\ 
 			& Pr(NM$|$Class=NM) & 0.994 & \\ 
 \hline
 		& Pr(Abs$|$Class=Abs) & 0.937 & [0.914, 0.977] \\ 
Florida	& Pr(Vot$|$Class=Vot) & 0.970 & [0.917, 0.982] \\ 
 		& Pr(NM$|$Class=NM) & 0.998 & \\ 
 \hline
 		& Pr(Abs$|$Class=Abs) & 0.928 & [0.915, 0.980] \\ 
 Kansas	& Pr(Vot$|$Class=Vot) & 0.953 & [0.918, 0.982] \\ 
 		& Pr(NM$|$Class=NM) & 0.996 &  \\ 
 \hline
 			& Pr(Abs$|$Class=Abs) & 0.907 & [0.921, 0.978] \\ 
 Kentucky		& Pr(Vot$|$Class=Vot) & 0.974 & [0.921, 0.978] \\ 
 			& Pr(NM$|$Class=NM) & 0.993 & \\ 
 \hline
 		 & Pr(Abs$|$Class=Abs) & 0.979 & [0.915, 0.986] \\ 
 Missouri	 & Pr(Vot$|$Class=Vot) & 0.947 & [0.904, 0.982] \\ 
  		 & Pr(NM$|$Class=NM) & 0.999 &  \\ 
 \hline
  		 	& Pr(Abs$|$Class=Abs) & 0.945 & [0.908, 0.991] \\ 
 New Jersey 	& Pr(Vot$|$Class=Vot) & 1.000 & [0.895, 0.987] \\ 
		 	& Pr(NM$|$Class=NM) & 0.999 &  \\ 
 \hline
  		 	& Pr(Abs$|$Class=Abs) & 0.970 & [0.917, 0.982] \\ 
 New York 	& Pr(Vot$|$Class=Vot) & 0.947 & [0.908, 0.985] \\ 
 			 & Pr(NM$|$Class=NM) & 0.987 &  \\ 
 \hline
	 	& Pr(Abs$|$Class=Abs) & 0.941 & [0.911, 0.985] \\ 
 Nevada	 & Pr(Vot$|$Class=Vot) & 0.963 & [0.915, 0.982] \\ 
 		 & Pr(NM$|$Class=NM) & 0.996 & \\ 

 \hline
			& Pr(Abs$|$Class=Abs) & 0.950 & [0.914, 0.986] \\ 
 Oklahoma	& Pr(Vot$|$Class=Vot) & 0.940 & [0.920, 0.980] \\ 
 			& Pr(NM$|$Class=NM) & 0.998 &  \\ 
  \hline
			 & Pr(Abs$|$Class=Abs) & 0.975 & [0.912, 0.981] \\ 
Pennsylvania	& Pr(Vot$|$Class=Vot) & 0.971 & [0.914, 0.986] \\ 
 			& Pr(NM$|$Class=NM) & 0.994 &  \\ 
  \hline
 				& Pr(Abs$|$Class=Abs) & 0.972 & [0.908, 0.979] \\ 
 Rhode Island 	  	& Pr(Vot$|$Class=Vot) & 0.953 & [0.912, 0.980] \\ 
  				& Pr(NM$|$Class=NM) & 0.997 & \\ 
  \hline
\end{tabular}
\end{center}
\caption{Yahtzee\texttrademark classifier results from 1000 randomly selected Facebook users from each state. Each user was given a classification based on the Yahtzee\texttrademark process. ``Abs'' $=$ Abstainer, ``Vot'' $=$ Voter, ``NM'' $=$ Not Matched. The conditional probabilities are calculated as the probability of observing a true behavior conditional on the Yahtzee\texttrademark classification. The 95\% confidence intervals are for the null distribution of 95\% accuracy in the classification, calculated from a binomial distribution with the same number of draws in each category.  In total, 22 of the 26 tests fall within these intervals, suggesting that deviations from 95\% accuracy are due to sampling variation, and for a large sample the procedure will generate the desired level of accuracy.}
\label{truth}
\end{table}
\clearpage

\clearpage
\bibliographystyle{plain}
\bibliography{YahtzeeMatchBib}

\begin{thebibliography}{10}

\bibitem{Bondetal:2011}
Robert~M. Bond, Christopher~J. Fariss, Jason~J. Jones, Adam Kramer, Jaime~E.
  Settle, Cameron Marlow, and James~H. Fowler.
\newblock A massive scale experiment in social influence and political
  mobilization.
\newblock {\em working paper}, 2011.

\bibitem{Cioffi-Revilla:2010}
Claudio Cioffi-Revilla.
\newblock Computational social science.
\newblock {\em Computational Statistics}, 2:259--271, 2010.

\bibitem{FreedmanNissimPinkas:2004}
Michael~J. Feedman, Kobbi Nissim, and Benny Pinkas.
\newblock Efficient private matching and set intersection.
\newblock In {\em Eurocrypt '2004 Proceedings, LNCS 3027, Springer-Verlag},
  pages 1--19, 2004.

\bibitem{Ginsbergetal:2009}
Jeremy Ginsberg, Matthew~H. Mohebbi, Rajan~S. Patel, Lynnette Brammer, Mark~S.
  Smolinski, and Larry Brilliant.
\newblock Detecting influenza epidemics using search engine query data.
\newblock {\em Nature}, 457(7232):1012--1014, 2009.

\bibitem{GolderMacy:2011}
Scott~A. Golder and Michael~W. Macy.
\newblock Diurnal and seasonal mood vary with work, sleep, and daylength across
  diverse cultures.
\newblock {\em Science}, 333(6051):878--1881, 2011.

\bibitem{Goldreich:2003}
Oded Goldreich.
\newblock Cryptography and cryptographic protocols.
\newblock {\em Distributed Computation}, 16:177--199, 2003.

\bibitem{GonzalezHidalgoBarabasi:2008}
M.~C. Gonzalez, C.~A. Hidalgo, and A.~L. Barab{\'a}si.
\newblock Understanding individual human mobility patterns.
\newblock {\em Nature}, 453:779--782, 2008.

\bibitem{Lazeretal:2009}
David Lazer, Alex Pentland, Lada Adamic, Sinan Aral, Albert-L{\'a}szl{\'o}
  Barab{\'a}si, Devon Brewer, Nicholas~A. Christakis, Noshir Contractor,
  James~H. Fowler, Myron Gutmann, Tony Jebara, Gary King, Michael Macy, Deb
  Roy, and Marshall {Van Alstyne}.
\newblock Computational social science.
\newblock {\em Science}, 323:721--723, 2009.

\bibitem{LindellPinkas:2009}
Yehuda Lindell and Benny Pinkas.
\newblock Secure multiparty computation for privacy-preserving data mining.
\newblock {\em Journal of Privacy and Confidentiality}, 1(1):59--98, 2009.

\bibitem{Misloveetal:2011}
Alan Mislove, Sune Lehmann, Yong-Yeol Ahn, Jukka-Pekka Onnela, and J.~Niels
  Rosenquist.
\newblock Pulse of the nation: U.s. mood throughout the day inferred from
  twitter.
\newblock {\em working paper}, 2011.

\bibitem{OnnelaaReed-Tsochas:2010}
Jukka-Pekka Onnelaa and Felix Reed-Tsochas.
\newblock Spontaneous emergence of social influence in online systems.
\newblock {\em Proceedings of the National Academy of Sciences},
  107(43):18375--18380, 2010.

\end{thebibliography}
\end{document}